\documentclass[10pt,conference]{IEEEtran}
\IEEEoverridecommandlockouts
\usepackage{cite}
\usepackage{amsmath,amssymb,amsfonts}
\usepackage{algorithmic}
\usepackage{graphicx}
\usepackage{textcomp}
\usepackage{xcolor, colortbl}
\usepackage{booktabs}
\usepackage{multirow}
\usepackage{xspace}
\usepackage{caption}
\usepackage{subcaption}
\usepackage{footnote}
\usepackage{url}
\usepackage{tcolorbox}
\usepackage[para]{footmisc}
\usepackage{array,ragged2e}
\usepackage{multicol}
\usepackage{listings}
\usepackage{balance}

\definecolor{Gray}{gray}{0.9}

\def\BibTeX{{\rm B\kern-.05em{\sc i\kern-.025em b}\kern-.08em
    T\kern-.1667em\lower.7ex\hbox{E}\kern-.125emX}}
\begin{document}

\title{Evaluating SZZ Implementations Through a Developer-informed Oracle}

\author{
\IEEEauthorblockN{Giovanni Rosa\IEEEauthorrefmark{1}, 
Luca Pascarella\IEEEauthorrefmark{2}, 
Simone Scalabrino\IEEEauthorrefmark{1}, 
Rosalia Tufano\IEEEauthorrefmark{2}, 
Gabriele Bavota\IEEEauthorrefmark{2},\\
Michele Lanza\IEEEauthorrefmark{2}, and 
Rocco Oliveto\IEEEauthorrefmark{1}}
\IEEEauthorblockA{
	\IEEEauthorrefmark{1}\emph{University of Molise, Italy}\\
	\IEEEauthorrefmark{2}\emph{Software Institute @ USI Università della Svizzera italiana, Switzerland}\\
}}

\maketitle

\newcommand{\ie}{\emph{i.e.,}\xspace}
\newcommand{\eg}{\emph{e.g.,}\xspace}
\newcommand{\etc}{etc.\xspace}
\newcommand{\etal}{\emph{et~al.}\xspace}
\newcommand{\secref}[1]{Section~\ref{#1}\xspace}
\newcommand{\figref}[1]{Fig.~\ref{#1}\xspace}
\newcommand{\listref}[1]{Listing~\ref{#1}\xspace}
\newcommand{\tabref}[1]{Table~\ref{#1}\xspace}
\newcommand{\tool}[1]{{\sc #1}\xspace}
\newcommand{\commitref}[1]{$\mathtt{#1}$\xspace}

\newcommand{\totalCommits}{19,603,736\xspace}
\newcommand{\validated}{3,585\xspace}
\newcommand{\finalInstances}{1,930\xspace}
\newcommand{\finalInstancesPercent}{55.6\%\xspace}
\newcommand{\finalInstancesIssues}{212\xspace}
\newcommand{\finalInstancesIssuesPercent}{11.0\%\xspace}
\newcommand{\oracle}{1,115\xspace}
\newcommand{\oracleWithIssue}{129\xspace}

\newboolean{showcomments}

\setboolean{showcomments}{true}

\ifthenelse{\boolean{showcomments}}
  {\newcommand{\nb}[2]{
    \fbox{\bfseries\sffamily\scriptsize#1}
    {\sf\small$\blacktriangleright$\textit{#2}$\blacktriangleleft$}
   }
  }
  {\newcommand{\nb}[2]{}
  }

\newcommand\ROCCO[1]{\textcolor{red}{\nb{ROCCO}{#1}}}
\newcommand\SIMONE[1]{\textcolor{red}{\nb{SIMONE}{#1}}}
\newcommand\LUCA[1]{\textcolor{red}{\nb{LUCA}{#1}}}
\newcommand\GIOVANNI[1]{\textcolor{red}{\nb{GIOVANNI}{#1}}}
\newcommand\ROSALIA[1]{\textcolor{red}{\nb{ROSALIA}{#1}}}
\newcommand\GABRIELE[1]{\textcolor{red}{\nb{GABRIELE}{#1}}}
\newcommand\MICHELE[1]{\textcolor{red}{\nb{MICHELE}{#1}}}

\newcommand{\bszz}{\textsc{B-SZZ}\xspace}
\newcommand{\agszz}{\textsc{AG-SZZ}\xspace}
\newcommand{\djszz}{\textsc{DJ-SZZ}\xspace}
\newcommand{\rszz}{\textsc{R-SZZ}\xspace}
\newcommand{\lszz}{\textsc{L-SZZ}\xspace}
\newcommand{\maszz}{\textsc{MA-SZZ}\xspace}
\newcommand{\raszz}{\textsc{RA-SZZ}\xspace}
\newcommand{\raszzs}{\textsc{RA-SZZ\textsuperscript{*}}\xspace}

\newcommand{\justcitebszz} {\cite{sliwerski2005changes}\xspace}
\newcommand{\justciteagszz}{\cite{kim2006automatic}\xspace}
\newcommand{\justcitedjszz}{\cite{williams2008szz}\xspace}
\newcommand{\justciterlszz}{\cite{davies2014comparing}\xspace}
\newcommand{\justcitemaszz}{\cite{da2016framework}\xspace}
\newcommand{\justciteraszz}{\cite{neto2018impact}\xspace}
\newcommand{\justciteraszzs}{\cite{neto2019revisiting}\xspace}

\newcommand{\citeallszz}{\cite{sliwerski2005changes, kim2006automatic, williams2008szz, davies2014comparing, da2016framework, neto2018impact, neto2019revisiting}\xspace}
\newcommand{\citeallimprovedszz}{\cite{kim2006automatic, williams2008szz, davies2014comparing, da2016framework, neto2018impact, neto2019revisiting}\xspace}

\newcommand{\citebszz}{\'Sliwerski \etal \cite{sliwerski2005changes}\xspace}
\newcommand{\citeagszz}{Kim \etal \cite{kim2006automatic}\xspace}
\newcommand{\citedjszz}{Williams and Spacco \cite{williams2008szz}\xspace}
\newcommand{\citerlszz}{Davies \etal \cite{davies2014comparing}\xspace}
\newcommand{\citemaszz}{da Costa \etal \cite{da2016framework}\xspace}
\newcommand{\citeraszz}{Neto \etal \cite{neto2018impact}\xspace}
\newcommand{\citeraszzs}{Neto \etal \cite{neto2019revisiting}\xspace}

\begin{abstract}

The SZZ algorithm for identifying bug-inducing changes has been widely used to evaluate defect prediction techniques and to empirically investigate when, how, and by whom bugs are introduced. Over the years, researchers have proposed several heuristics to improve the SZZ accuracy, providing various implementations of SZZ. 
However, fairly evaluating those implementations on a reliable oracle is an open problem: SZZ evaluations usually rely on (i) the manual analysis of the SZZ output to classify the identified bug-inducing commits as true or false positives; or (ii) a golden set linking bug-fixing and bug-inducing commits. In both cases, these manual evaluations are performed by researchers with limited knowledge of the studied subject systems. Ideally, there should be a golden set created by the original developers of the studied systems. 

We propose a methodology to build a ``developer-informed'' oracle for the evaluation of SZZ variants. We use Natural Language Processing (NLP) to identify bug-fixing commits in which developers explicitly reference the commit(s) that introduced a fixed bug. This was followed by a manual filtering step aimed at ensuring the quality and accuracy of the oracle. Once built, we used the oracle to evaluate several variants of the SZZ algorithm in terms of their accuracy. Our evaluation helped us to distill a set of  lessons learned to further improve the SZZ algorithm.

\end{abstract}



\newcommand{\manual}{Manually defined (researchers)\xspace}
\newcommand{\bszzuses}{\cite{palomba2018diffuseness, pascarella2019fine, ccaglayan2016effect, wen2016locus, posnett2013dual, kim2008classifying, tan2015online, kononenko2015investigating, wehaibi2016examining, lenarduzzi2020sonarqube}}
\newcommand{\agszzuses}{\cite{tufano2017empirical, bernardi2018relation, hata2012bug, rahman2011bugcache, eyolfson2014correlations, misirli2016studying, canfora2011long, prechelt2014software, bird2009fair}}
\newcommand{\djszzuses}{\cite{marinescu2014covrig, borg2019szz, bavota2015four, toth2016public, fan2019impact, karampatsis2019often, rodriguez2020bugs, rodriguez2018reproducibility}}
\newcommand{\rlszzuses}{\cite{da2016framework}}
\newcommand{\maszzuses}{\cite{fan2019impact, neto2018impact, neto2019revisiting, tu2020better, aman2019empirical, chen2019extracting}}
\newcommand{\raszzuses}{\cite{fan2019impact, neto2018impact, yan2020just}}
\newcommand{\raszzsuses}{None}

\section{Introduction} \label{sec:intro}

The SZZ algorithm, proposed by \'Sliwerski, Zimmermann, and Zeller \cite{sliwerski2005changes} at MSR 2005, identifies, given a bug-fixing commit $C_{BF}$, the commits that likely introduced the bug fixed in $C_{BF}$. These commits are termed ``bug-inducing'' commits. In essence, given $C_{BF}$ as input, SZZ identifies the last change (commit) to each source code line changed in $C_{BF}$ (\ie changed to fix the bug). This is done by relying on the annotation/blame feature of versioning systems. The identified commits are considered as the ones that later on triggered the bug-fixing commit $C_{BF}$. 

SZZ has been widely adopted to (i) design and evaluate defect prediction techniques \cite{hata2012bug, tan2015online, pascarella2019fine, yan2020just, fan2019impact}, and to (ii) run empirical studies aimed at investigating under which circumstances bugs are introduced \cite{bavota2015four, tufano2017empirical, aman2019empirical, chen2019extracting}. The relevance of the SZZ algorithm was recognized a decade later with a MIP (Most Influential Paper award) presented at the 12th Working Conference on Mining Software Repositories (MSR 2015).

Several researchers have proposed variants of the original algorithm, with the goal of boosting its accuracy \citeallimprovedszz. 

For example, one issue with the basic SZZ implementation is that it considers changes to code comments and whitespaces like any other change. 

This means that if a comment is modified in $C_{BF}$, the latest change to that comment is mistakenly considered as a bug-inducing commit. An improvement by Kim \etal \cite{kim2006automatic} was therefore to ignore changes to code comments and blank lines as candidate bug-inducing commits. 

\begin{table*}
 \centering
 \resizebox{\linewidth}{!}{
 \begin{tabular}{l|l l l| l r r}
  \toprule
  \textbf{Approach name}         & \textbf{Reference}                      & \textbf{Based on}                  & \textbf{Used by} & \textbf{Oracle type}           & \textbf{\# Projects}  & \textbf{\# Bug Fixes} \\
  \midrule                                                                                                                         
  \bszz                          & \citebszz                               &                                    & \bszzuses        & //                             & //                 & //              \\
  \agszz                         & \citeagszz                              & \bszz                              & \agszzuses       & \manual                        & 2                  & 301             \\
  \djszz                         & \citedjszz                              & \agszz                             & \djszzuses       & \manual                        & 1                  & 25              \\
  \lszz \& \rszz                 & \citerlszz                              & \agszz                             & \rlszzuses       & \manual                        & 3                  & 174             \\
  \maszz                         & \citemaszz                              & \agszz                             & \maszzuses       & Automatically computed metrics & 10                 & 2,637           \\
  \raszz                         & \citeraszz                              & \maszz                             & \raszzuses       & \manual                        & 10                 & 365             \\
  \raszzs                        & \citeraszzs                             & \raszz                             & \raszzsuses      & \manual                        & 10                 & 365             \\
  \bottomrule
\end{tabular}
}
 \vspace{-0.1cm}
 \caption{Variants of the SZZ algorithm. For each one, we specify (i) the algorithm on which it is based, (ii) references of works using it, (iii) the oracle used in the evaluation (how it was built, number of projects and bug fixes considered).}
 \label{tab:approaches}
\end{table*}

Despite the major advances made on the accuracy of SZZ, Alencar da Costa \etal \cite{da2016framework} highlighted the major difficulties in fairly evaluating and comparing the SZZ variants proposed in the literature. They observed that the studies presenting and evaluating SZZ variants mostly rely on manual analysis of a small sample of SZZ results \cite{sliwerski2005changes,kim2006automatic,williams2008szz,davies2014comparing}, with the goal of evaluating its accuracy. Such an evaluation is usually performed by the researchers who---not being the original developers of the studied systems---do not always have the knowledge needed to correctly identify the bug introducing commit. Also, due to the high cost of such a manual analysis, it is usually performed on a small sample of the identified bug-inducing commits. Other researchers built instead a ground truth to evaluate the performance of the SZZ algorithm \cite{neto2019revisiting}. However, also in these cases, the ground truth is produced by the researchers. Alencar da Costa \etal \cite{da2016framework} called for evaluations performed with ``\emph{domain experts (\eg developers or testers)}'' reporting however that ``\emph{such an analysis is impractical}'' since ``\emph{the experts would need to verify a large sample of bug-introducing changes, which is difficult to scale up to the size of modern defect datasets}'' \cite{da2016framework}.

We present a methodology to build a ``developer-informed'' oracle for the evaluation of SZZ implementations. To explain its idea, let us take as example commit \commitref{a8a97bd} from the {\tt apache/thrift} GitHub project, accompanied by a commit message saying: ``\emph{THRIFT-4513: fix bug in comparator introduced by e58f75d}''. The developer fixing the bug is explicitly documenting the commit that introduced such a bug. Based on this observation, we defined a number of strict NLP-based heuristics to automatically identify notes in bug-fixing commits in which developers explicitly reference the commit(s) that introduced the fixed bug. We applied these heuristics to a total of \totalCommits mined through GH Archive \cite{githubarchive}, which archives all public events on GitHub. 

Our goal with the above described process is not to be exhaustive, \ie we do not want to identify all bug-fixing commits in which developers indicated the bug-inducing commit(s), but rather to obtain a high-quality dataset of commits that were certainly of the bug-inducing kind.

We mined the time period between March 2011 and April 2020, obtaining \validated commits. To further  increase the intrinsic quality of the dataset, we manually validated the \validated commits, to (i) verify if, from the commit message, it was clear that the developer was documenting the bug-inducing commit; and (ii) taking note of any issue referenced in the commit message (\eg issue THRIFT-4513 in the previous example). Information from the issue tracker is exploited by some of the SZZ implementations and we wanted our dataset to include it.

As output of this process, we obtained a dataset of \finalInstances validated bug-fixing commits in which developers documented the commit(s) that introduced the bug, with \finalInstancesIssues also including information about the fixed issue(s). To the best of our knowledge, our work is the first presenting a dataset for the SZZ evaluation built by using information about the bug-inducing commit(s) explicitly reported by the bug fixer. 

We tested nine variants of SZZ on our dataset. Besides reporting their precision and recall, we analyzed their complementarity and focused on the set of bug-fixes where all SZZ variants fail. A qualitative analysis of those cases allowed to distill lessons learned useful to further improve the SZZ algorithm in the future. Summarizing, our contributions are:

\begin{enumerate}

\item A methodology to build a ``developer-informed'' oracle for the evaluation of SZZ implementations, which does not require major manual efforts as compared to the classical manual identification of bug-inducing commits.

\item A first, easily extensible dataset built using our methodology and featuring \finalInstances validated bug-fixing commits.

\item An empirical study comparing the effectiveness of several SZZ implementations.

\item A comprehensive replication package featuring (i) the dataset, and (ii) the implemented SZZ variants \cite{replication}.

\end{enumerate}

 
\section{Background and Related Work} \label{sec:related}

We start by presenting several variants of the SZZ algorithm \cite{sliwerski2005changes} proposed in the literature over the years. Then, we discuss how those variants have been used in SE research community.

\subsection{SZZ and its variants}

\tabref{tab:approaches} presents the SZZ variants proposed in the literature.
We report for each of them its name and reference, the approach it builds upon (\ie the starting point on which the authors provide improvements), some references to works that used it, and information about the oracle used for the evaluation. Specifically, we report how the oracle was built and the number of projects/bug reports considered. 

All the approaches that aim at identifying bug-inducing commits (BICs) rely on two elements: (i) the revision history of the software project, and (ii) an issue tracking system (optional, needed only by some SZZ implementations).

The original SZZ algorithm was proposed by \citebszz (we refer to it as \bszz, following the notation provided by da Costa \etal \cite{da2016framework}). \bszz takes as input a bug report from an issue tracking system, and tries to find the commit that fixes the bug. To do this, \bszz uses a two-level confidence level: \textit{syntactic} (possible references to the bug ID in the issue tracker) and \textit{semantic} (\eg the bug description is contained in the commit message). \bszz relies on the CVS \texttt{diff} command to detect the lines changed in the fix commit and the \texttt{annotate} command to find the commits in which the lines were modified. Using this procedure, \bszz determines the \textit{earlier} change at the location of the fix. Potential bug-inducing commits performed after the bug was reported are always ignored.

\citeagszz noticed that \bszz has limitations mostly related to formatting/cosmetic changes (\eg moving a bracket to the next line). Such changes can deceive \bszz: \bszz (i) can report as BIC a revision which only changed the code formatting, and (ii) it can consider as part of a bug-fix a formatting change unrelated to the actual fix. They introduce a variant (\agszz) in which they used an annotation graph, a data structure associating the modified lines with the containing function/method. \agszz also ignores the cosmetic parts of the bug-fixes to provide more precise results.

\citedjszz improved the \agszz algorithm in two ways: first, they use a line-number mapping approach \cite{williams2008branching} instead of the annotation graph introduced by \citeagszz; second, they use DiffJ \cite{pace2007tool}, a Java syntax-aware diff tool, which allows their approach (which we call \djszz) to exclude non-executable changes (\eg \texttt{import} statements).

\citerlszz propose two variations on the criterion used to select the BIC among the candidates: \lszz uses the largest candidate, while \rszz uses the latest one. These improvements were done on top of the \agszz algorithm.

\maszz, introduced by \citemaszz, excludes from the candidate BICs all the \textit{meta-changes}, \ie commits that do not change the source code. This includes (i) branch changes, which are copy operations from one branch to another, (ii) merge changes, which consist in applying the changes performed in a branch to another one, and (iii) property changes, which only modify file properties (\eg permissions).

To further reduce the false positives, two new variants were introduced by Neto \etal, \raszz \justciteraszz and \raszzs \justciteraszzs. Both exclude from the BIC candidates the refactoring operations, \ie changes that should not modify the behavior of the program. Both approaches use state-of-the-art tools: \raszz uses RefDiff \cite{silva2017refdiff}, while \raszzs uses Refactoring Miner \cite{tsantalis2018accurate}, with the second one being more effective \justciteraszzs.

The original SZZ was not empirically evaluated \justcitebszz. Instead, all its variants, except \maszz, were manually evaluated by their authors. One of them, \raszzs \justciteraszzs, used an external dataset, \ie Defect4J \cite{just2014defects4j}. \maszz was evaluated using automated metrics, namely \textit{earliest bug appearance}, \textit{future impact of a change}, and \textit{realism of bug introduction} \cite{da2016framework}.

In \tabref{tab:tools} we list the open-source implementations of SZZ. 

\begin{table}[ht]
 \centering
 \resizebox{\linewidth}{!}{
 \begin{tabular}{l l l}
  \toprule
  \textbf{Tool name}                       & \textbf{Approach}                      & \textbf{Public repository}                            \\
  \midrule
  SZZ Unleashed \cite{borg2019szz}         & $\sim$\djszz \cite{williams2008szz}      & \url{https://github.com/wogscpar/SZZUnleashed} \\
  OpenSZZ \cite{lenarduzzi2020openszz}     & $\sim$\bszz  \cite{sliwerski2005changes} & \url{https://github.com/clowee/OpenSZZ}        \\
  \textsc{PyDriller} \cite{Spadini2018}    & $\sim$\agszz \cite{sliwerski2005changes}  & \url{https://github.com/ishepard/pydriller} \\
  \bottomrule
 \end{tabular}
 }
 \caption{Open-source tools implementing SZZ.}
 \label{tab:tools}
 
\end{table}

SZZ Unleashed \cite{borg2019szz} partially implements \djszz: it uses line-number mapping \cite{williams2008szz} but it does not rely on DiffJ \cite{pace2007tool} for computing diffs, also working on non-Java files. It does not take into account meta-changes \justcitemaszz and refactorings \cite{neto2019revisiting}.

OpenSZZ \cite{lenarduzzi2020openszz} implements the basic version of the approach, \bszz. Since it is based on the git \texttt{blame} command, it implicitly uses the annotated graph \cite{kim2006automatic}. 

\textsc{PyDriller} \cite{Spadini2018}, a general purpose tool for analyzing git repositories, also implements B-SZZ. It uses a simple heuristic for ignoring C- and Python-style comment lines, as proposed by \citeagszz. We do not report in \tabref{tab:tools} a comprehensive list of all the SZZ implementations that can be found on GitHub, but only the ones presented in papers.

\subsection{SZZ in Software Engineering Research}

The original SZZ algorithm and its variations were used in a plethora of studies. We discuss some examples, while for a complete list we refer to the extensive literature review by Rodr\'iguez-P\'erez \etal \cite{rodriguez2018reproducibility}, featuring 187 papers.

SZZ has been used to run several empirical investigations having different goals \cite{ccaglayan2016effect,lenarduzzi2020sonarqube,wehaibi2016examining,tufano2017empirical,bernardi2018relation,eyolfson2014correlations,misirli2016studying,canfora2011long,prechelt2014software,bird2009fair,rodriguez2018reproducibility,aman2019empirical,chen2019extracting,posnett2013dual,karampatsis2019often,bavota2015four,kononenko2015investigating,palomba2018diffuseness}. For example, Aman \etal \cite{aman2019empirical} studied the role of local variable names in fault-introducing commits and they used SZZ to retrieve such commits, while Palomba \etal \cite{palomba2018diffuseness} focused on the  impact of code smells, and used SZZ to determine whether an artifact was smelly when a fault was introduced. Many studies also leverage SZZ to evaluate defect prediction approaches \cite{kim2008classifying,tan2015online,hata2012bug,rahman2011bugcache,toth2016public,tu2020better,wen2016locus,yan2020just,fan2019impact,pascarella2019fine}. 

Looking at \tabref{tab:approaches} it is worth noting that, despite its clear limitations \cite{kim2006automatic}, many studies, even recent ones, still rely on \bszz \bszzuses{} (the approaches that use git implicitly use the annotation graph defined by Kim \etal \cite{kim2006automatic}). Improvements are only slowly adopted in the literature, possibly due to the fact that some of them are not released as tools and that the two standalone tools providing a public SZZ implementation were released only recently \cite{lenarduzzi2020openszz, borg2019szz}.

The studies most similar to ours are the one by da Costa \etal \cite{da2016framework} and the one by Rodr{\'\i}guez-P{\'e}rez \etal \cite{rodriguez2020bugs}. Both report a comparison of different SZZ variants. Da Costa \etal \cite{da2016framework} defined and used a set of metrics for evaluating SZZ implementations without relying on a manually defined oracle. However, they specify that, ideally, domain experts should be involved in the construction of the dataset \cite{da2016framework}, which motivated our study.
Rodr{\'\i}guez-P{\'e}rez \etal \cite{rodriguez2018reproducibility} introduced a model for distinguishing bugs caused by modifications to the source code (the ones that SZZ algorithms can detect) and the ones that are introduced due to problems with external dependencies. They also used the model to define a manually curated dataset on which they evaluated SZZ variants. Their dataset is created by researchers and not domain experts. In our study, instead, we rely on the explicit information provided by domain experts in their commit messages.

 
\section{Building a Developer-informed Dataset of Bug-inducing Commits} \label{sec:dataset}

We present a methodology to build a dataset of bug-inducing commits by exploiting information provided by developers when fixing bugs. Our methodology reduces the manual effort required for building such a dataset and more important, does not assume technical knowledge of the involved source code on the researchers' side.

\begin{figure*}
 \includegraphics[width=\linewidth]{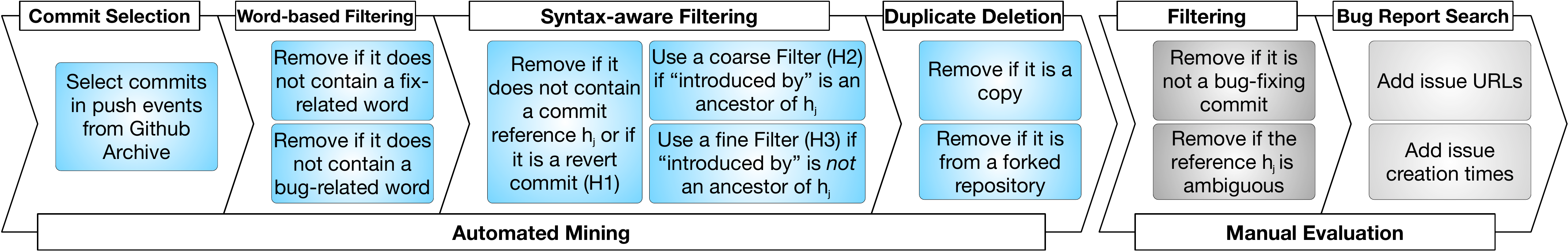}
 \caption{Process used for building the dataset.}
 \label{fig:dataset}
\end{figure*}

The proposed methodology involves two main steps: (i) automatic mining from open-source repositories of bug-fixing commits in which developers explicitly indicate the commit(s) that introduced the fixed bug, and (ii) a manual filtering aimed at improving the dataset quality by removing ambiguous commit messages that do not give confidence in the information provided by the developer. In the following, we detail these two steps. The whole process is depicted in \figref{fig:dataset}.

\subsection{Mining Bug-fixing and Bug-inducing Commits}

There are two main approaches proposed in the literature for selecting bug-fixing commits. The first one relies on the linking between commits and issues \cite{linking}: issues labeled with ``bug'', ``defect'', \etc are mined from the issue tracking system, storing their issue ID (\eg THRIFT-4513). Then, commits referencing the issue ID are mined from the versioning system and identified as bug-fixing commit. While such a heuristic is fairly precise, it has two important drawbacks that make it unsuitable for our work. First, the link to the issue tracking system must be known and a specific crawler for each different type of issue tracker (\eg Jira, Bugzilla, GitHub, \etc) must be built.

Second, projects can use a customized set of labels to indicate bug-related issues. Manually extracting this information for a large set of repositories is expensive. The basic idea behind this first phase is to use the commit messages to identify bug-fixing commits: we automatically analyze bug-fixing commit messages searching for those explicitly referencing bug-inducing commits. 

As a preliminary step, we mined \textsc{GH Archive} \cite{githubarchive} which provides, on a regular basis, a snapshot of public events generated on GitHub in the form of JSON files. 

We mined the time period going from March 1$^{st}$ 2011 to April 30$^{th}$ 2020, extracting \totalCommits commits performed in the context of \textit{push} events: such events gather the commits done by a developer on a repository before performing the \textit{push} action. Considering the goal of building an oracle for SZZ algorithms, we are not interested in any specific programming language.
We performed three steps to select a candidate set of commits to manually analyze in the second phase: (i) we selected a first candidate set of bug-fixing commits, (ii) we used syntax-aware heuristics to refine such a set, and (iii) we removed duplicates.

\subsubsection{Word-Based Bug-Fixing Selection}

To identify bug-fixing commits, we first apply a lightweight regular expression on all the commits we gathered, as done in previous work \cite{Fischer:ICSM2003,Tufano:tosem2019}. We mark as potential bug-fixes all commits accompanied by a message including at least a fix-related word\footnote{\emph{fix} or \emph{solve}\label{fnote:fixwords}} and a bug-related word\footnote{\emph{bug}, \emph{issue}, \emph{problem}, \emph{error}, or \emph{misfeature}\label{fnote:bugwords}}. We exclude the messages that include the word \emph{merge} to ignore merge commits. Note that we do not need such a heuristic to be 100\% precise, since two additional and more precise steps will be performed on the identified set of candidate fixing commits to exclude false positives (\ie a NLP-based step and a manual analysis). 

\subsubsection{Syntax-Aware Filtering}

We needed to select from the set of candidate bug-fixing commits only the ones in which developers likely documented the bug-inducing commit(s). We used the syntax-aware heuristics described below to do this. The first author defined such heuristics through a trial-and-error procedure, taking a 1-month time period of events on GH Archive to test and refine different versions of the heuristics, manually inspecting the achieved results after each run. The final version has been consolidated with the feedback of two additional authors.

As a preliminary step, we used the \texttt{doc.sents} function of the \textsc{spaCy}\footnote{https://spacy.io/} Python module for NLP to extract the set $S_c$ of sentences composing each commit message $c$. 

\begin{figure}[ht]
	\centering
	\includegraphics[width=0.65\linewidth]{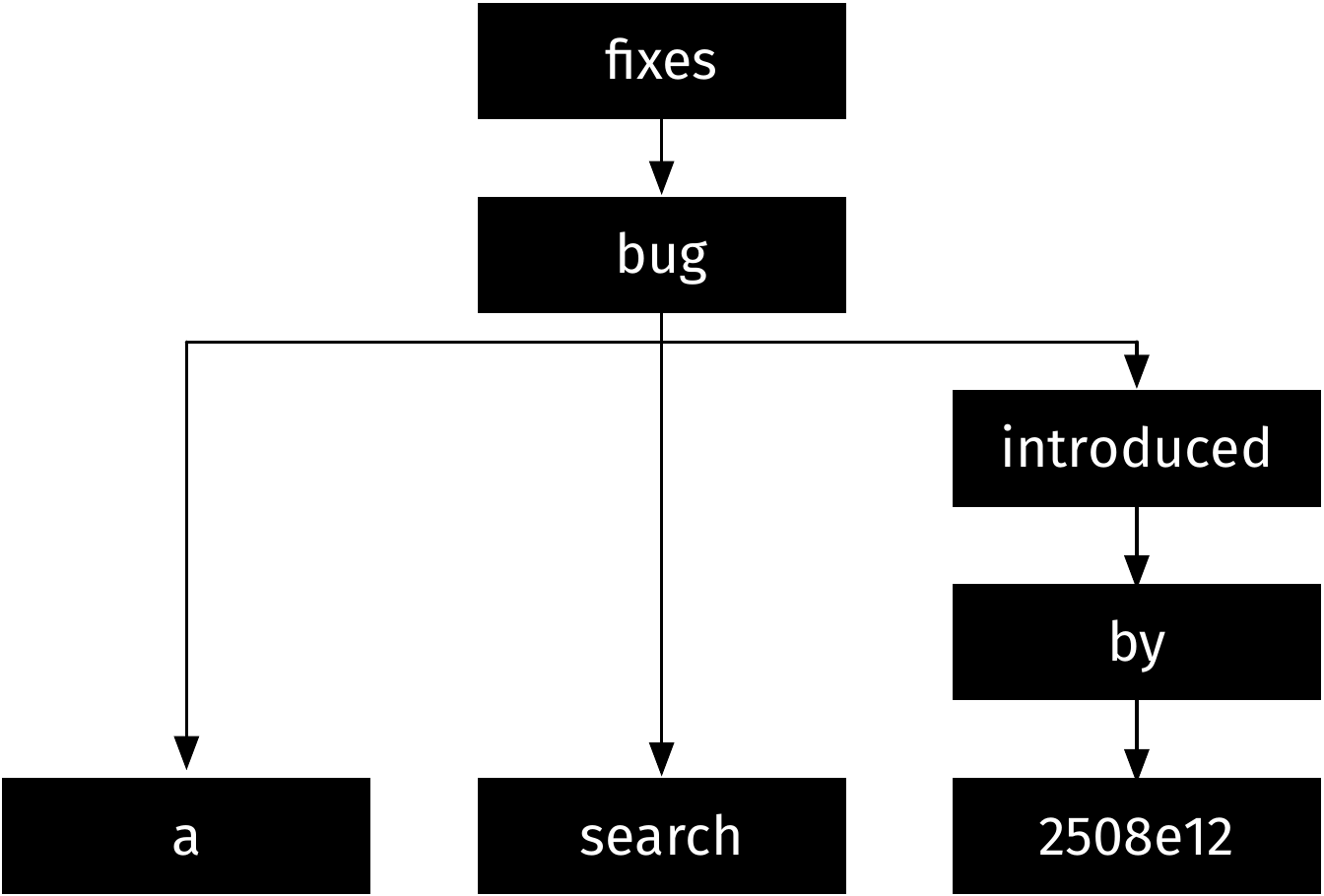}
	\caption{Example of word dependency tree built by \textsc{spaCy}.}
	\label{fig:dependency}
	\vspace{-0.3cm}
\end{figure}

For each sentence $s_i \in S_c$, we used \textsc{spaCy} to build its word dependency tree $t_i$, \ie a tree containing the syntactic relationships between the words composing the sentence.
\figref{fig:dependency} provides an example of $t_i$ generated for the sentence ``\emph{fixes a search bug introduced by 2508e12}''.

By navigating the word dependency tree, we can infer that the verb ``fix'' refers to the noun ``bug'', and that the verb ``introduced'' is linked to commit id \commitref{2508e12} through the ``by'' apposition. 

\textbf{H1: Exclude Commits Without Reference and Reverts.} We split each $s_i \in S_c$ into words and we select all its commit hashes $H(s_i)$ using a regular expression\footnote{[0-9a-f]\{6,40\}}. We ignore all the $s_i$ for which $H(s_i)$ is empty (\ie which do not mention any commit hash). Similarly, we filter out all the $s_i$ that either (i) start with a commit hash, or (ii) include the verb ``revert'' referring to any $h_j \in H(s_i)$. We keep all the remaining $s_i$. We exclude the commits that do not contain any valid sentence as for this heuristic. We use the $H(s_i)$ extracted with this heuristic also for the following heuristics.

\textbf{H2: Coarsely Filter Explicit Introducing References.} If one of the ancestors of $h_j$ is the verb ``introduce'' (in any declension), as it happens in \figref{fig:dependency}, we consider this as a strong indication of the fact that the developer is indicating $h_j$ as (one of) the bug-inducing commit(s). In this case, we check if $h_j$ also includes at least one of the fix-related words\footref{fnote:fixwords} \textbf{and} one of the bug-related words\footref{fnote:bugwords} as one of its ancestors or children. At least one of the two words (\ie the one indicating the fixing activity or the one referring to a bug) must be an ancestor. We do this to avoid erroneously selecting sentences such as ``\emph{Improving feature introduced in 2508e12 and fixed a bug}'', in which both the fix-related and the bug-related word are children of $h_j$. 

For example, the $h_j$ in \figref{fig:dependency} meets this constraint since it has among its ancestors both \emph{fix} and \emph{bug}. We also exclude the cases in which the words \emph{attempt} or \emph{test} (again, in different declensions) appear as ancestors of $h_j$. We do this to exclude false positives observed while experimenting with earlier versions of this heuristic. 

For example, the sentence ``\emph{Remove attempt to fix error introduced in 2f780609}'' belongs to a commit that aims at reverting previous changes. Similarly, the sentence ``\emph{Add tests for the fix of the bug introduced in 2f780609}'' most likely belongs to the message of a test-introduction commit. 

\textbf{H3: Finely Filter Non-Explicit Introducing References.} If $h_j$ does not contain the verb ``introduce'' as one of its ancestors, we apply a finer filtering heuristic: both a word indicating a fixing activity \textbf{and} a word indicating a bug must appear as one of $h_j$'s ancestors. Also, we define a list of stop-words that must not appear either in the $h_j$'s ancestor as well as in the dependencies (\ie ancestors and children) of the ``fixing activity'' word. Such a stop-word list, derived through a trial-and-error procedure, includes eight additional words ({\textit{was}, \textit{been}, \textit{seem}, \textit{solved}, \textit{fixed}, \textit{try}, \textit{trie} (to capture \textit{tries} and \textit{tried}), and \textit{by}), besides \textit{attempt} and \textit{test} also used in H2. This allows, for example, to exclude sentences such as ``\emph{This definitely fixes the bug I tried to fix in commit 26f3fe2}'', meets all selection criteria for H3 but it is a false positive.

\subsubsection{Duplicate Deletion}

We saved the list of commits including at least one sentence $s_i$ meeting H1 and either H2 or H3 in a MySQL database. Since we analyzed a large set of projects, it was frequent that some commits were duplicated due to the fact that different forks of a given project are available. As a final step, we removed such duplicates, keeping only the commit of the main project repository.

Out of the \totalCommits parsed commits, the automated filtering selected \validated commits. Our goal with the above described process is not to be exhaustive, \ie we do not want to identify all bug-fixing commits in which developers indicated the bug-inducing commit(s), but rather to obtain a high-quality dataset of commits that were certainly of the bug-inducing kind. The quality of the dataset is then further increased during the subsequent step of manual analysis.

\subsection{Manual Analysis}

Four of the authors (from now on, evaluators) manually inspected the \validated commits produced by the previous step. The  evaluators have different backgrounds (graduate student, faculty member, junior and a senior researcher with two years of industrial experience). The goal of the manual validation was to verify (i) whether the commit was an actual bug-fix, and (ii) if it included in the commit message a non-ambiguous sentence clearly indicating the commit(s) in which the fixed bug was introduced. For both steps the evaluators mostly relied on the commit message and, if available, on possible references to the issue tracker. Those references could be issue IDs or links that the evaluators inspected to (i) ensure that the fixed issue was a bug, and (ii) store for each commit the links to the mentioned issues and, for each issue, its opening date. 

The latter is an information that may be required by an SZZ implementation (\eg SZZ Unleashed \cite{borg2019szz} and OpenSZZ \cite{lenarduzzi2020openszz} require the link to the issue) to exclude from the candidate list of bug-inducing commits those performed after the opening of the fixed issue. 

Indeed, if the fixed bug has been already reported at date $d_i$, a commit performed on date $d_j > d_i$ cannot be responsible for its introduction. Since the commits to inspect come from a variety of software systems, they rely on different issue trackers. When an explicit link was not available but an issue was mentioned in the commit message (\eg see the commit message shown in the introduction), the evaluators searched for the project's issue tracker, looking on the GitHub repository for documentation pointing to it (in case the project did not use the GitHub issue tracker itself). If no information was found, an additional Google search was performed, looking for the project website or directly searching for the issue ID mentioned in the commit message.

The manual validation was supported by a web-based application we developed that assigns to each evaluator the candidate commits to review, showing for each of them its commit message and a clickable link to the commit \textsc{GitHub} page. Using a form, the evaluator indicated whether the commit was relevant for the oracle (\ie an actual bug-fix documenting the bug-inducing commit) or not, and listing mentioned issues together with their opening date. Each commit was assigned by the web application to two different evaluators, for a total of 7,170 evaluations. To be more conservative and to have higher confidence in our oracle, we decided to not resolve conflicts (\ie cases in which one evaluator marked the commit as relevant and the other as irrelevant): we excluded from our oracle all commits with at least one ``irrelevant'' flag.

\subsection{The Obtained SZZ Oracles}

Out of the \validated manually validated commits, \finalInstances (\finalInstancesPercent) passed our manual filtering, of which  \finalInstancesIssues include references to a valid issue (\ie an issue labeled as a bug that can be found online). This indicates that SZZ implementations that rely on information from issue trackers can only be run on a minority of bug-fixing commits. Indeed, the \finalInstances instances we report have been manually checked as true positive bug-fixes, and only \finalInstancesIssues of these (\finalInstancesIssuesPercent) mention the fixed issue. The dataset is available in our replication package \cite{replication}.

These \finalInstances commits and their related bug-inducing commits impact files written in many different languages. All the implementations of the SZZ algorithm (except for B-SZZ) perform some language-specific parsing to ignore changes performed to code comments. 

In our study (\secref{sec:design}) we experimented several versions of the SZZ including those requiring the parsing of comments. We implemented support for the top-8 programming languages present in our oracle (\ie the ones responsible for more code commits): C, C++, C\#, Java, JavaScript, Ruby, PHP, and Python. This led to the creation of the dataset we use in our experimentation, only including bug-fixing/inducing commits impacting files written in one of the eight programming languages we support. This dataset is also available in our replication package \cite{replication}. \tabref{tab:datasetinfo} summarizes the main characteristics of the \emph{overall} dataset and of the \emph{language-filtered} one. 

\begin{table}[ht]
 \centering
 \resizebox{\linewidth}{!}{
 \begin{tabular}{l |r r r| r r r}
  \toprule
                             & \multicolumn{3}{c|}{\textit{Overall}}                               & \multicolumn{3}{c}{\textit{Language-filtered}}                    \\
  \textbf{Language}          & \textbf{\#Repos} & \textbf{\#Commits} & \textbf{\#Issues} & \textbf{\#Repos} & \textbf{\#Commits} & \textbf{\#Issues} \\
  \midrule                                                                                        
  C                          &              350 &                433 &                52 &              297 &                366 &               41  \\
  Python                     &              271 &                304 &                36 &              249 &                279 &               35  \\
  C++                        &              198 &                241 &                31 &              138 &                162 &               20  \\
  JS                         &              169 &                180 &                26 &              127 &                135 &               18  \\
  Java                       &               88 &                101 &                14 &               72 &                 80 &               10  \\
  PHP                        &               63 &                 71 &                 6 &               56 &                 64 &                5  \\
  Ruby                       &               43 &                 47 &                 5 &               36 &                 37 &                4  \\
  C\#                        &               25 &                 32 &                 3 &               20 &                 27 &                1  \\
  Others                     &              498 &                588 &                48 &                0 &                  0 &                0  \\
  \midrule
  \textbf{Total}             &            1,625 &              1,930 &               212 &              951 &              1,115 &              129  \\
  \bottomrule
 \end{tabular}
  }
 \caption{Features of the \textit{language-filtered}/\textit{overall} datasets.}
 \label{tab:datasetinfo}
\end{table}

It is worth noting that a repository or even a commit can involve several programming languages: for this reason, the \textit{total} may be lower than the sum of the per-language values (\ie a repository can be counted in two or more languages).

Besides sharing the datasets as JSON files, we also share the cloned repositories from which the bug-fixing commits have been extracted. This enables the replication of our study and the use of the datasets for the assessment of future SZZ improvements.

\section{Study Design} \label{sec:design}

\newcommand\rqOne{How do different variants of SZZ perform in identifying bug-inducing changes?}
\newcommand{\oracleAllDataset}{$\mathit{oracle}_{\mathit{all}}$\xspace}
\newcommand{\oracleIssueDataset}{$\mathit{oracle}_{\mathit{issues}}$\space}
\newcommand{\oracleAllDatasetJava}{$\mathit{oracle}_{\mathit{all}}^{J}$\xspace}
\newcommand{\oracleIssueDatasetJava}{$\mathit{oracle}_{\mathit{issues}}^{J}$\space}

\newlength\Linewidth
\def\findlength{
	\setlength\Linewidth\linewidth
	\addtolength\Linewidth{-4\fboxrule}
	\addtolength\Linewidth{-3\fboxsep}
}

\newenvironment{researchquestion}
	{\par\begingroup
	\setlength{\fboxsep}{5pt}\findlength
	\setbox0=\vbox\bgroup\noindent
	\hsize=0.95\linewidth
	\begin{minipage}{0.95\linewidth}\normalsize}
	{\end{minipage}\egroup
	   \vspace{3pt}
	\textcolor{gray}{\fboxsep1.5pt\fbox{\fboxsep5pt\colorbox{white}{\normalcolor\box0}}}
	\endgroup\par\noindent
	\normalcolor\ignorespacesafterend}

The \emph{goal} of this study is to experiment several implementations of the SZZ algorithm on the previously defined \emph{language-filtered} dataset (\emph{context} of our study). The \emph{perspective} is that of researchers interested in assessing the effectiveness of the state-of-the-art implementations and identify possible improvements that can be implemented to further improve the accuracy of the SZZ algorithm. To achieve such a goal, we aim to answer the following research question:

\newcolumntype{R}[1]{>{\RaggedRight\arraybackslash}p{#1}}
\begin{table*}[ht]
	\centering
	\caption{Characteristics of the SZZ implementations we compare in our study. We mark with a ``$\diamond$'' our re-implementations.\vspace{0cm}}
	\label{tab:experiment}%
	\resizebox{\linewidth}{!}{
	\begin{tabular}{R{5em} R{10em} R{12em} R{8em} R{6em} p{21em}}
		\toprule
		\textbf{Acronym} & \textbf{Fix Line Filtering}                                & \textbf{BIC Identification Method}  & \textbf{BIC Filtering}                                & \textbf{BIC Selection}        & \textbf{Differences w.r.t. the original paper} \\
		\midrule                                                                                                    
		\bszz            & //                                                         & Annotation Graph\justciteagszz      & //                                                    & //                            & We use git \texttt{blame} instead of the CVS \texttt{annotate}, \ie we implicitly use an annotation graph \justciteagszz. We do not filter BICs based on the issue creation date.\textsuperscript{$\diamond$} \\
		\midrule                                                                                                    
		\agszz           & Cosmetic changes\justciteagszz                             & Annotation Graph\justciteagszz      & //                                                    & //                            & No differences.\textsuperscript{$\diamond$} \\
		\midrule          
		\maszz           & Cosmetic changes\justciteagszz                             & Annotation Graph\justciteagszz      & Meta-Changes\justcitemaszz                            & //                            & No differences.\textsuperscript{$\diamond$} \\
		\midrule          
		\lszz            & Cosmetic Changes\justciteagszz                             & Annotation Graph\justciteagszz      & Meta-Changes\justcitemaszz                            & Largest \justciterlszz        & We filter meta-changes \justcitemaszz.\textsuperscript{$\diamond$} \\
		\midrule          
		\rszz            & Cosmetic Changes\justciteagszz                             & Annotation Graph\justciteagszz      & Meta-Changes\justcitemaszz                            & Latest \justciterlszz         & We filter meta-changes \justcitemaszz.\textsuperscript{$\diamond$} \\
		\midrule          
		\raszzs          & Cosmetic Changes\justciteagszz Refactorings\justciteraszzs & Annotation Graph\justciteagszz      & Meta-Changes\justcitemaszz                            & //                            & We use Refactoring Miner 2.0 \cite{Tsantalis:TSE:2020:RefactoringMiner2.0}.\textsuperscript{$\diamond$} \\
		\midrule          
		SZZ@PYD          & Cosmetic Changes\justciteagszz                             & Annotation Graph\justciteagszz      & //                                                    & //                            & We implement a wrapper for \textsc{PyDriller} \cite{Spadini2018}.  \\
		\midrule          
		SZZ@UNL          & Cosmetic Changes\justciteagszz                             & Line-number Mapping\justcitedjszz   & Issue-date\justcitebszz                               & //                            & We implement a wrapper for SZZ Unleashed \cite{borg2019szz}.   \\
		\midrule          
		SZZ@OPN          & //                                                         & Annotation Graph\justciteagszz      & //                                                    & //                            & We implement a wrapper for OpenSZZ \cite{lenarduzzi2020openszz}. \\
		\bottomrule
	\end{tabular}
	}
	\vspace{-0.3cm}
\end{table*}

\begin{center}	
	\begin{researchquestion}
		\emph{\rqOne}
	\end{researchquestion}	 
\end{center}

\subsection{Data Collection}

We focused our experiment on several variants of the SZZ algorithm. Specifically, we (i) re-implemented all the main approaches available in the literature (presented in \secref{sec:related}) in a new tool, and (ii) adapted three existing tools (\textsc{PyDriller} \cite{Spadini2018}, SZZ Unleashed \cite{borg2019szz}, and OpenSZZ \cite{lenarduzzi2020openszz}) to work with our dataset. We provide in our replication package \cite{replication} both our tool and the adapted versions of the other tools, including detailed instructions on how to run them.

We report the details about all the implementations we compare in \tabref{tab:experiment} and, for each of them, we explicitly mention (i) how it filters the lines changed in the fix (\eg it removes cosmetic changes), (ii) which methodology it uses for identifying the preliminary set of bug-inducing commits (\eg annotation graph), (iii) how it filters such a preliminary set (\eg it removes meta-changes), and (iv) if it uses a heuristic for selecting a single bug-inducing commit and, if so, which one (\eg most recent commit). 

We also explicitly mention any difference between our implementations and the approaches as described in the original papers presenting them. 

It is worth noting that we intentionally made all our re-implementations optionally independent from the issue-tracker systems: we did this because most of the instances of our dataset do not provide links to the bug-report ($\sim$88\%). This is the reason why we did not implement the ``Issue-date'' as a BIC filtering technique by default, despite it is reported in the respective papers (\eg for \bszz). However, we experiment all techniques with and without such a filtering applied. 

As for the tools, instead, we did not modify their implementation of the BIC-finding procedures: \eg we did not remove the filtering by issue date from SZZ Unleashed. On the other hand, we implemented wrappers for such tools that allowed us to run them with our dataset. SZZ Unleashed depends on a specific issue-tracker system (\ie Jira) for filtering commits done after the bug-report was opened. We made it independent from it by adapting our datasets to the input it expects (\ie Jira issues in JSON format). It is worth noting that, despite the complexity of such files, SZZ Unleashed only uses the issue opening date in its implementation. For this reason, we only provide such field and we set the others to \texttt{null}.

Note that some of the original implementations listed in \tabref{tab:experiment} can identify bug-fixing commits. In our study, we did not want to test such a feature: we test a scenario in which the implementations already have the bug-fixing commits for which they should detect the bug-inducing commit(s).

To evaluate the previously described implementations, we defined two datasets extracted from the \emph{language-filtered} dataset: (i) the \oracleAllDataset dataset, featuring \oracle bug-fixes, which includes both the ones with and without issue information, and (ii) the \oracleIssueDataset dataset, featuring \oracleWithIssue instances, which includes only instances with issue information.
Also, we defined two additional datasets, \oracleAllDatasetJava (80 instances) and \oracleIssueDatasetJava (10 instances), obtained by considering only Java-related commits from the \oracleAllDataset and \oracleIssueDataset, respectively. We did this because two implementations, \ie \raszzs\footnote{It relies on Refactoring Miner \cite{Tsantalis:TSE:2020:RefactoringMiner2.0} which only works on Java files.} and OpenSZZ, only work on Java files.

We ran all the implementations on all the datasets on which they can be executed (\ie we did not run \raszzs and OpenSZZ on the datasets including non-Java files). It is worth noting that SZZ Unleashed requires the issue date in order to work, so it would not be possible to run it on the \oracleAllDataset dataset. To avoid this problem, we simulated the best-case-scenario for such commits: we pretended that an issue about the bug was created few seconds after the last bug-inducing commit was done. Consider the bug-fixing commit $BF$ without issue information and its set of bug-inducing commits $BIC$; we assumed that the issue mentioned in $BF$ had $\mathit{max_{b \in BIC}}(date(b)) + \delta$ as opening date, where $\delta$ is a small time interval (we used 60 seconds). 


\subsection{Data Analysis}

Given the defined oracle and the set of bug-inducing commits detected by the experimented implementations, we evaluated its accuracy by using two widely-adopted Information Retrieval (IR) metrics, namely recall and precision \cite{Baeza-Yates:1999}. 

We computed them using the following formulas:

\scriptsize
\vspace{-0.4cm}
\begin{multicols}{2}
  $$
    \mathit{recall} = {{|\mathit{correct} \cap \mathit{identified}|} \over {|\mathit{correct}|}}
  $$\break
  $$
    \mathit{precision} = {{|\mathit{correct} \cap \mathit{identified}|} \over {|\mathit{identified}|}}
  $$
\end{multicols}
\normalsize

\noindent where $\mathit{correct}$ and $\mathit{identified}$ represent the set of true positive bug-inducing commits (those indicated by the developers in the commit message) and the set of bug-inducing commits detected by the experimented algorithm, respectively. As an aggregate indicator of precision and recall, we report the F-measure \cite{Baeza-Yates:1999}, defined as the harmonic mean of precision and recall. Such metrics were also used in previous work for evaluating SZZ variants (\eg \citeraszzs).

Given the set of experimented SZZ variants/tools $SZZ_{exp}=\{v_1, v_2, \dots v_n\}$, we also analyze their complementarity by computing the following metrics for each $v_i$ \cite{oliveto:icpc2010}:

{\scriptsize
$$
\mathit{correct}_{v_i \cap v_j} = {|\mathit{correct}_{v_i} \cap \mathit{correct}_{v_j}| \over |\mathit{correct}_{v_i} \cup \mathit{correct}_{v_j}|}
$$
$$
\mathit{correct}_{v_i \setminus (\mathit{SZZ}_{\mathit{exp}} \setminus v_i)} = {|\mathit{correct}_{v_i} \setminus \mathit{correct}_{(\mathit{SZZ}_{\mathit{exp}} \setminus v_i)}| \over |\mathit{correct}_{v_i} \cup \mathit{correct}_{(\mathit{SZZ}_{\mathit{exp}} \setminus v_i}|}
$$
}
\noindent where $\mathit{correct}_{v_i}$ represents the set of correct bug-inducing commits detected by $v_i$ and $\mathit{correct}_{(SZZ_{exp} \setminus v_i)}$ the correct bug-inducing commits detected by all other techniques but $v_i$. $\mathit{correct}_{v_i \cap v_j}$ measures the overlap between the set of correct bug-inducing commits identified by two given implementations: we computed it between all the pairs of SZZ variants and present the results using a heatmap. $\mathit{correct}_{v_i \setminus (\mathit{SZZ}_{\mathit{exp}} \setminus v_i)}$, instead, measures the correct bug-inducing commits identified only by technique $v_i$ and missed by all others. 

It is worth clarifying that, when we compute the overlap metrics, we compare all the implementations among them on the same dataset. This means, for example, that we do not compute the overlap between a variant tested on \oracleAllDataset and another variant tested on \oracleIssueDataset.

As a last step in our analysis, we compute the set of bug-fixing commits for which none of the experimented techniques was able to correctly identify the bug-inducing commit(s). We qualitatively discuss these cases to understand their peculiarities and point to future improvements of the SZZ algorithm.

\section{Results Discussion} \label{sec:results}

\tabref{tab:prec-recall-no-date} reports the results achieved by the experimented SZZ variants and tools.


\begin{table}[ht]
	\centering
	\caption{Precision, recall, and F-measure calculated for all SZZ algorithms. $\dagger$ Java only files.\vspace{0cm}}
	\scriptsize
	\begin{tabular}{l|lccc|ccc}
		\toprule

		& \multirow{2}{*}{Algorithm} & \multicolumn{3}{c|}{\emph{oracle$_{all}$}}  & \multicolumn{3}{c}{\emph{oracle$_{issue}$}} \\

		& & \multicolumn{1}{c}{\textbf{Recall}} & \multicolumn{1}{c}{\textbf{Precision}} & \multicolumn{1}{c|}{\textbf{F1}} & \multicolumn{1}{c}{\textbf{Recall}} & \multicolumn{1}{c}{\textbf{Precision}} & \multicolumn{1}{c}{\textbf{F1}} \\
		\midrule
		
		\multirow{9}{*}{\rotatebox[origin=c]{90}{No issue date filter}} & \bszz      		&  0.69  &  0.39  &  0.50  &  0.69  &  0.38  &  0.49  \\
																		& \agszz     		&  0.60  &  0.45  &  0.52  &  0.62  &  0.43  &  0.51  \\
																		& \lszz      		&  0.45  &  0.52  &  0.48  &  0.43  &  0.49  &  0.46  \\
																		& \rszz      		&  0.57  &  0.66  &  0.61  &  0.56  &  0.64  &  0.60  \\
																		& \maszz     		&  0.64  &  0.36  &  0.46  &  0.65  &  0.36  &  0.47  \\
																		& $\dagger$\raszzs  &  0.45  &  0.35  &  0.39  &  0.40  &  0.57  &  0.47  \\
																		& SZZ@PYD    		&  0.67  &  0.39  &  0.49  &  0.68  &  0.39  &  0.50  \\
																		& SZZ@UNL    		&  0.72  &  0.09  &  0.16  &  0.72  &  0.06  &  0.12  \\
																		& $\dagger$SZZ@OPN  &  0.19  &  0.32  &  0.24  &  0.10  &  0.50  &  0.17  \\
		\midrule

		\multirow{9}{*}{\rotatebox[origin=c]{90}{With date filter}} 	& \bszz      		&  0.69  &  0.42  &  0.53  &  0.69  &  0.39  &  0.50  \\
																		& \agszz     		&  0.60  &  0.49  &  0.54  &  0.62  &  0.44  &  0.52  \\
																		& \lszz      		&  0.45  &  0.54  &  0.49  &  0.43  &  0.50  &  0.46  \\
																		& \rszz      		&  0.57  &  0.73  &  0.64  &  0.56  &  0.67  &  0.61  \\
																		& \maszz     		&  0.64  &  0.39  &  0.48  &  0.65  &  0.37  &  0.47  \\
																		& $\dagger$\raszzs  &  0.45  &  0.43  &  0.44  &  0.40  &  0.57  &  0.47  \\
																		& SZZ@PYD    		&  0.67  &  0.42  &  0.52  &  0.68  &  0.41  &  0.51  \\
																		& SZZ@UNL    		&  0.72  &  0.09  &  0.16  &  0.72  &  0.06  &  0.12  \\
																		& $\dagger$SZZ@OPN  &  0.19  &  0.33  &  0.24  &  0.10  &  0.50  &  0.17  \\
		\bottomrule
	\end{tabular}%
	\label{tab:prec-recall-no-date}%
	\vspace{-0.3cm}
\end{table}

The top part of the table shows the results when the issue date filter has not been applied, while the bottom part relates to the application of the date filter. With ``issue date filter'' we refer to the process through which we remove from the list of candidate bug-inducing commits returned by a given technique all those performed after the issue documenting the bug has been opened. Those are known to be false positives. For this reason, such a filter is expected to not have any impact on recall (since the discarded bug-inducing commits should all be false positives) while increasing precision. The left part of \tabref{tab:prec-recall-no-date} shows the results achieved on $oracle_{all}$, while the right part focuses on $oracle_{issue}$. 

The first result to extrapolate from \tabref{tab:prec-recall-no-date} is the general trend concerning the performance of the SZZ implementations.

When not using the issue date filtering (top part), the highest achieved F-Measure is 61\% (\rszz). \rszz uses the annotation graph, ignores cosmetic changes and meta-changes, and only considers as bug-inducing commits the latest change that impacted a line changed to fix the bug. 

Such a combination of heuristics make the \rszz the most precise on both oracles, achieving a 66\% precision on $oracle_{all}$ and 64\% on $oracle_{issue}$. With respect to recall/precision tradeoff, there is a price to pay in terms of recall that, however, it is not dramatically worse compared to the best approach in terms of recall: SZZ@UNL (SZZ Unleashed). The latter achieves a 72\% recall on both $oracle_{all}$ and $oracle_{issue}$ datasets, with, however, a precision of 9\% and 6\%, respectively. We investigated the reasons behind such a low precision, finding that it is mainly due to a set of outlier bug-fixing commits for which SZZ@UNL identifies a high number of (false positive) bug-inducing commits. For example, three bug-fixing commits are responsible for 72 identified bug-inducing commits, out of which only three are correct. We analyzed the distribution of bug-inducing commits reported by SZZ@UNL for the different bug-fixing commits. Cases for which more than 20 bug-inducing commits are identified for a single bug-fix can be considered outliers. By ignoring those cases, the recall and precision of SZZ@UNL are 67\% and 18\%, respectively on $oracle_{all}$, and 67\% and 17\% on $oracle_{issue}$. By lowering the outlier threshold to 10 bug-inducing, the precision grows in both datasets to 24\%. We believe that the low precision of SZZ@UNL may be due to misbehavior of the tool in few isolated cases.

Two implementations (\ie \raszzs and SZZ@OPN) only work with Java files. In this case, we compute their recall and precision assuming by only considering the bug-fixing commits impacting Java files. Both of them exhibit limited recall and precision. While this is due in part to limitations of the implementations, it is also worth noting that the number of Java-related commits in our datasets is quite limited (\ie 80 in $oracle_{all}$ and only 10 in $oracle_{issue}$). Thus, failing on a few of those cases penalizes in terms of performance metrics. Still, we found the low precision of \raszzs surprising, considering the expensive mechanism it uses to limit false positives (\ie ignoring lines impacted by refactoring operations detected by Refactoring Miner \cite{Tsantalis:TSE:2020:RefactoringMiner2.0}. 

\bszz, the simplest SZZ version, exhibits a good recall of 69\% on both datasets, making it the second-best algorithm after SZZ@UNL. Nonetheless, \bszz pays in precision, making it the fourth algorithm together with the PyDriller implementation for $oracle_{all}$ and the fifth for  $oracle_{issue}$. The similarity between \bszz and the PyDriller implementation results in very similar performances. 

\agszz, \lszz, and \maszz exhibit, as compared to others, good performance for both recall and precision. These algorithms provide a good balance between recall and precision, as also shown by their F-Measure ($\sim$50\%).

The bottom of \tabref{tab:prec-recall-no-date} shows the results achieved by the same algorithms when using the issue data filter.

As expected, the recall remains equal to the previous scenario in all cases, with marginal improvements in precision (thanks to the removal of some false positives). While most of the algorithms improve their precision by 1\%-4\%, two algorithms obtain substantial improvements in the $oracle_{all}$ dataset: \raszzs (+8\%) and \rszz (+7\%). 

This boosts the latter to a very good 73\% precision on $oracle_{all}$, and 67\% on $oracle_{issue}$ (+3\%).

To summarize the achieved results: We found that \rszz is the most precise variant on our datasets, with a precision $\sim$70\% when the issue date filter is applied. Thus, we recommend it when precision is more important than recall (\eg when a set of bug-inducing commits must be mined for qualitative analysis). SZZ@UNL ensures instead a high recall at, however, a high precision cost. If the focus is on recall, the current recommendation is to rely on \bszz, using, for example, the implementation provided by PyDriller. Finally, looking at \tabref{tab:prec-recall-no-date}, it is clear that there are still margins of improvement for the accuracy of the SZZ algorithm. We discuss possible directions for future work in \secref{sub:improvements}.

\begin{figure*}[t!]
    \centering
    \begin{subfigure}[t]{0.49\textwidth}
        \centering
        \includegraphics[width=\linewidth]{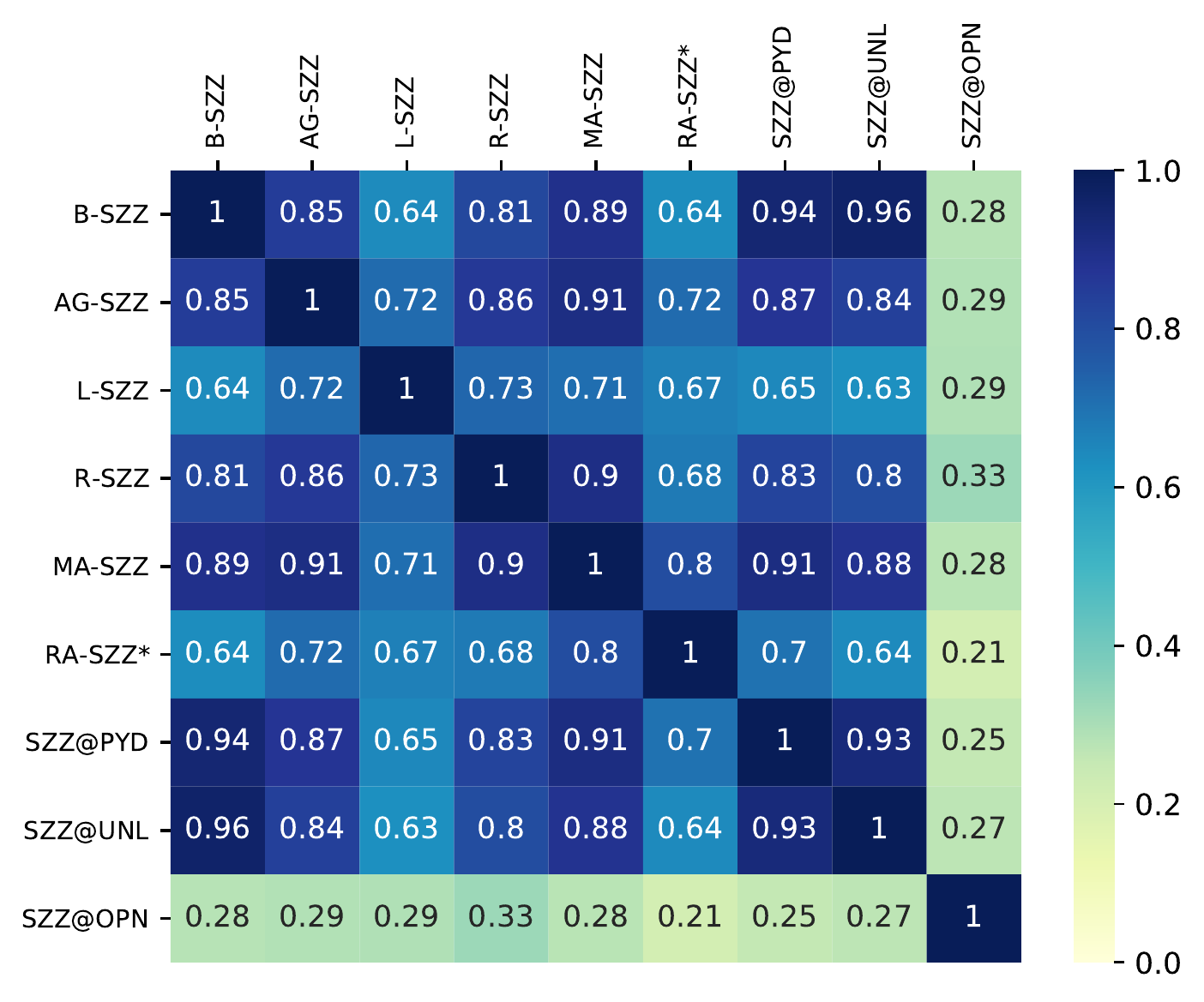}
		\caption{\emph{oracle$_{all}$}}
	\label{fig:heatmap-all}
    \end{subfigure}%
    ~ 
    \begin{subfigure}[t]{0.49\textwidth}
        \centering
        \includegraphics[width=\linewidth]{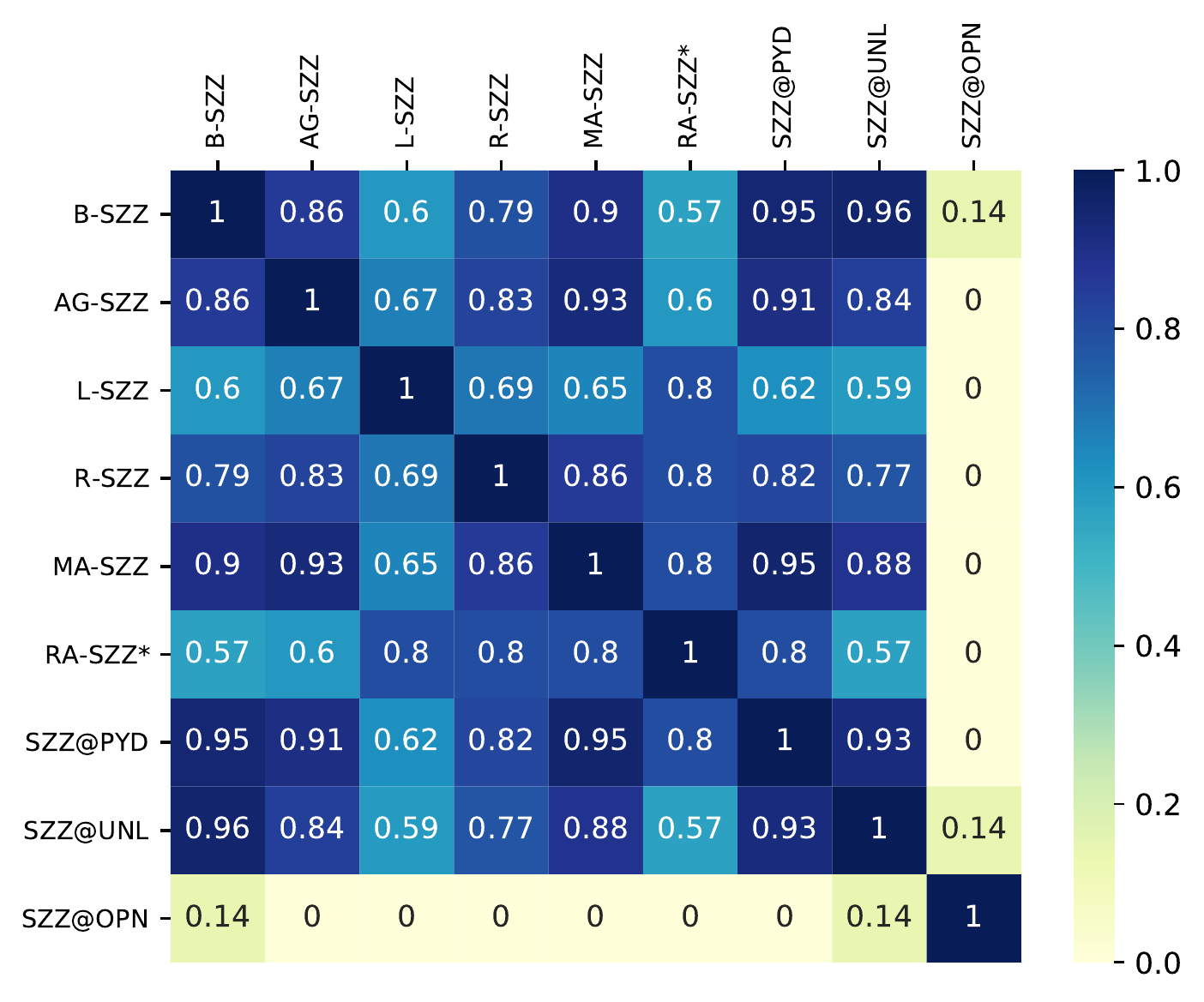}
	\caption{\emph{oracle$_{issue}$}}
	\label{fig:heat-map-issue-only}
    \end{subfigure}
    \caption{Overlap between SZZ variants when no issue date filter is applied.}
    \label{fig:heatmap}
\end{figure*}

\begin{figure*}[t!]
    \centering
    \begin{subfigure}[t]{0.49\textwidth}
        \centering
        \includegraphics[width=\linewidth]{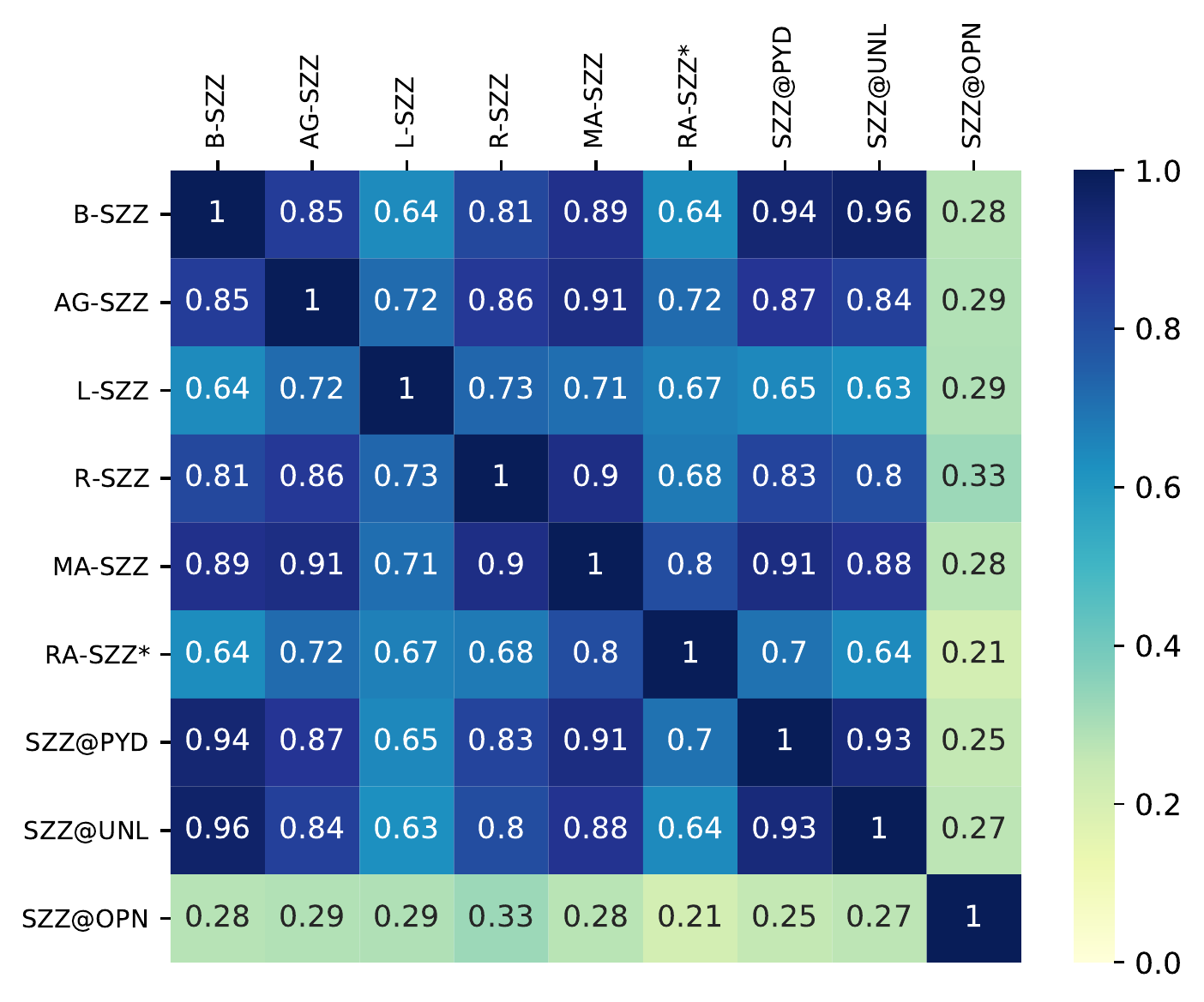}
		\caption{\emph{oracle$_{all}$}}
	\label{fig:heatmap-all-filter}
    \end{subfigure}%
    ~ 
    \begin{subfigure}[t]{0.49\textwidth}
        \centering
        \includegraphics[width=\linewidth]{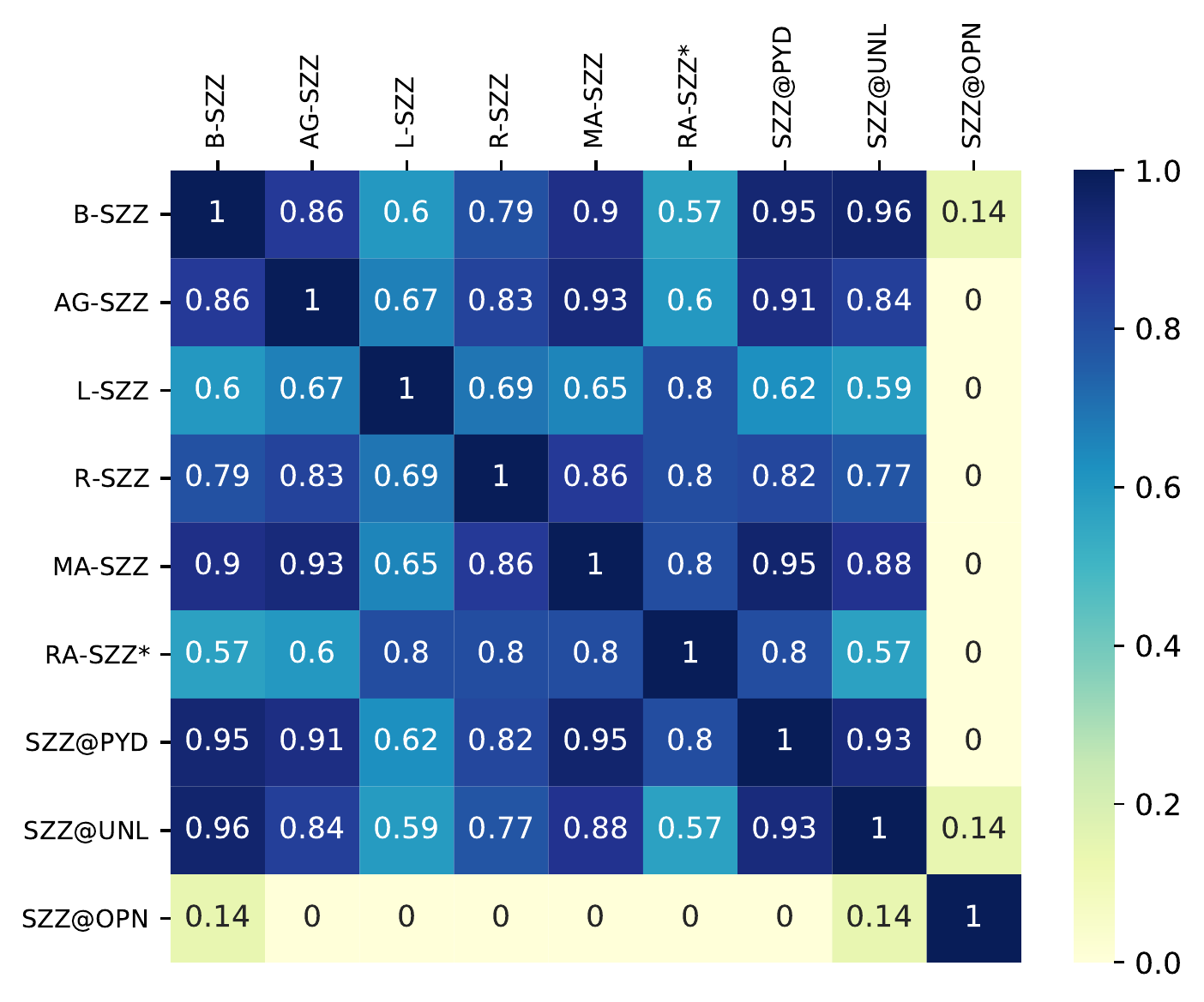}
	\caption{\emph{oracle$_{issue}$}}
	\label{fig:heat-map-issue-only-filter}
    \end{subfigure}
    \caption{Overlap between SZZ variants when the issue date filter is applied.}
    \label{fig:heatmap-filter}
\end{figure*}

\tabref{tab:overlap} shows the $\mathit{correct}_{v_i \setminus (\mathit{SZZ}_{\mathit{exp}} \setminus v_i)}$ metric we computed for each SZZ variant $v_i$. 


\begin{table}[ht]
	\centering
	\caption{Bug inducing commits correctly identified exclusively by the $v_i$ algorithm. $\dagger$ Java only files.}
	\scriptsize
	\begin{tabular}{lll|ll}
		\toprule

		\multirow{2}{*}{Algorithm} & \multicolumn{2}{c}{No date filter}  & \multicolumn{2}{c}{With date filter} \\

		                           & \emph{oracle$_{all}$}     & \emph{oracle$_{issue}$} & \emph{oracle$_{all}$} & \emph{oracle$_{issue}$} \\
		\midrule

		\bszz       		&  0/804         & 0/94         & 0/804          & 0/94         \\
		\agszz     		 	&  0/804         & 0/94         & 0/804          & 0/94         \\
		\lszz       		&  0/804         & 0/94         & 0/804          & 0/94         \\
		\rszz       		&  0/804         & 0/94         & 0/804          & 0/94         \\
		\maszz      		&  0/804         & 0/94         & 0/804          & 0/94         \\
		$\dagger$\raszzs    &  0/56          & 0/7          & 0/56           & 0/7          \\
		SZZ@PYD     		&  0/804         & 0/94         & 0/804          & 0/94         \\
		SZZ@UNL 	        & 20/804 (2.5\%) & 3/94 (3.2\%) & 20/804 (2.5\%) & 3/94 (2.2\%) \\
		$\dagger$SZZ@OPN    &  0/56          & 0/7          & 0/56           & 0/7          \\
		\bottomrule
	\end{tabular}%
	\label{tab:overlap}%
\end{table}

This metric measures the correct bug-inducing commits identified only by technique $v_i$ and missed by all the others.

\figref{fig:heatmap-all} and \figref{fig:heat-map-issue-only} depicts the $\mathit{correct}_{v_i \cap v_j}$ metric computed between each pair of SZZ variants when not filtering based on the issue date, while \figref{fig:heatmap-all-filter} and \figref{fig:heat-map-issue-only-filter} show the same metric when the issue filter has been applied. Given the metric definition, the depicted heatmaps are symmetric (\ie $\mathit{correct}_{v_i \cap v_j}$ = $\mathit{correct}_{v_j \cap v_i}$). The only technique able to identify bug-inducing commits missed by all others SZZ implementations is SZZ@UNL (20 on $oracle_{all}$ and 3 on $oracle_{issue}$) -- \tabref{tab:overlap}. This is not surprising considering the high SZZ@UNL recall and the high number of bug-inducing commits it returns for certain bug-fixes. It also explains why none of the other implementations identifies bug-inducing commits missed by all the others: Given 804 as cardinality of the intersection of the true positives identified by all SZZ techniques, SZZ@UNL correctly retrieves 800 of them.

Looking at the overlap metrics in \figref{fig:heatmap} and \figref{fig:heatmap-filter}, two observations can be made. First, the overlap in the identified true positives is substantial. Excluding SZZ@OPN, 21 of the 28 comparisons have an overlap in the identified true positives $\geq$70\% and the lower values are still in the range 60-70\%. The low overlap values observed for SZZ@OPN are instead due to the its low recall. Second, the complementarity between the different SZZ variants is quite low, which indicates that there is a set of bug-fixing commits for which all of the variants fail in identifying the correct bug-inducing commit(s).

We manually analyzed those cases to derive possible future improvements to the SZZ.

\subsection{Improvements to SZZ} \label{sub:improvements}

The manual analysis of 311 bug-fixing commits on which all SZZ variants fail allowed us to identify recurring patterns and distill three possible ways to improve the SZZ algorithm.

\subsubsection{The buggy line is not always impacted in the bug-fix} In some cases, fixing a bug introduced in line $l$ may not result in changes performed to $l$. An example in Java is the addition of an {\tt if} guard statement checking for {\tt null} values before accessing a variable. 

In this case, while the bug has been introduced with the code accessing the variable without checking whether it is {\tt null}, the bug-fixing commit does not impact such a line, it just adds the needed {\tt if} statement. An example from our dataset is the bug-fixing commit from the \emph{thcrap} repository \cite{github_adjacent_fix} in which line 289 is modified to fix a bug introduced in commit \commitref{b67116d}, as pointed by the developer in the commit message. However, the bug was introduced with changes performed on line 290 \cite{github_adjacent_fix}. Thus, running git blame on line 289 of the fix commit will retrieve a wrong bug-inducing commit. Defining approaches to identify the correct bug-inducing commit in these cases is far from trivial. 

However, by manually analyzing a large dataset of bug-fixing commits, it should be possible to identify fixing patterns with associated buggy lines. Such a dataset could be used to train a model able, given a bug-fixing commit, to point to the location of the bug.

\subsubsection{SZZ is sensible to history rewritings} Bird \etal \cite{bird2009promises} highlighted some of the peril of mining git repositories, among which the possibility for developers to rewrite the change history. This can be achieved through rebasing, for example: using such a strategy can have an impact on mining the change history \cite{kovalenko2018mining}, and, therefore, on the performance of the SZZ algorithm. Besides rebasing, git allows to partially rewrite history by reverting changes introduced in one or more commits in the past. This action is often performed by developers when a task they are working on leads to dead end. Once run, the revert command results in new commits in the change history that turn back the indicated changes. Consequently, SZZ can improperly show one of these commits as candidate bug-inducing.

For example, in the message of commit \commitref{5d8cee1} from the \emph{xkb-switch} project \cite{github_revert_fix}, the developer indicates that the bug she is fixing has been introduced in commit \commitref{42abcc}. By performing a blame on the fix commit, git returns as a bug-inducing commit \commitref{8b9cf29} \cite{github_revert_bic_wrong}, which is a revert commit. By performing an additional blame step, the correct bug-inducing commit pointed by the developer can be retrieved \cite{github_revert_bic_correct}. Future SZZ variants should handle revert commits, and properly deal with them when analyzing the change history.

\subsubsection{Looking at the ``big picture'' in code changes} In several bug-fixing commits we inspected, the implemented changes included both added and modified/deleted lines. SZZ implementations focus on the latter, since there is no way to blame a newly added line. However, we found cases in which the added lines were responsible for the bug-fixing, while the modified/deleted ones were unrelated. There have been a recent attempt to address this problem: Sahal and Tosun \cite{sahal2018identifying} proposed an SZZ extension that considers past history of all the lines in the block in which the added line appears. However, the research in this aspect is still at the beginning.

An example is commit \commitref{ca11949} from the \emph{snake} repository \cite{github_block_level_fix}, in which two lines are added and two deleted to fix a bug. The deleted lines, while being the target of SZZ, are unrelated to the bug-fix, as clear from the commit message pointing to commit \commitref{315a64b} \cite{github_block_level_bic} as the one responsible for the bug introduction. In the bug-inducing commit, the developer refactored the code to simplify an {\tt if} condition. While refactoring the code, she introduced a bug (\ie she missed an {\tt else} branch). The fixing adds the {\tt else} branch to the sequence of {\tt if}/{\tt else if} branches introduced in the bug-inducing commit. In this case, by relying on static analysis, it should be possible to link the added lines, representing the {\tt else} branch, to the set of {\tt if}/{\tt else if} statements preceding it. While the added lines cannot be blamed, lines related to them (\eg acting on the same variable, being part of the same ``high-level construct'' like in this case) could be blamed to increase the chances of identifying the bug-inducing commit. 

While this would help recall, it would penalize precision without careful heuristics aimed at filtering out false positives.

\section{Threats to Validity} \label{sec:threats}

{\em Construct validity.} During the manual validation, the evaluators mainly relied on the commit message and the linked issue(s), when available, to confirm that a mined commit was a bug-fixing commit. Misleading information in the commit message could result in the introduction of false positive instances in our dataset. However, all commits have been checked by at least two evaluators and doubtful cases have been excluded, privileging a conservative approach. 
To build our dataset, we considered all the projects from GitHub, without explicitly defining criteria to select only projects that are invested in software quality. Our assumption is that the fact that developers take care of documenting the bug-introducing commit(s) is an indication that they care about software quality. To ensure that the commits in our dataset are from projects that take quality into account, we manually analyzed 123 projects from our dataset, which allowed us to cover a significant sample of commits (286 out of 1,115, with 95\%$\pm$5\% confidence level). For each of them, we checked if they contained elements that indicate a certain degree of attention to software quality, \ie (i) unit test cases, (ii) code reviews (through pull requests), (iii) and continuous integration pipelines. We found that in 95\% of the projects, developers (i) wrote unit test cases, and (ii) conducted code reviews through pull requests. Also, we found CI pipelines in 75\% of the projects.

{\em Internal validity.} There is a possible subjectiveness introduced of the manual analysis, which has been mitigated with multiple evaluators per bug-fix. Also, we reimplemented most of the experimented SZZ approaches, thus possibly introducing variations as compared to what proposed by the original authors. We followed the description of the approaches in the original papers, documented in \tabref{tab:experiment} any difference between our implementations and the original proposals, and share our implementations \cite{replication}. Also, note that the differences documented in \tabref{tab:experiment} always aim at improving the performance of the SZZ variants and, thus, should not be detrimental for their performance.

{\em External validity.} While it is true that we mined millions of commits to build our dataset, we used very strict filtering criteria that resulted in \finalInstances instances for our oracle. Also, the SZZ implementations have been experimented on a smaller dataset of \oracle instances that is, however, still larger than those used in previous works. 
Finally, our dataset represents a subset of the bug-fixes performed by developers. This is due to our design choice, where we used strict selection criteria when building our oracle to prefer quality over quantity. It is possible that our dataset is biased towards a specific type of bug-fixing commits: there might be an inherent difference between the bug fixes for which developers document the bug-inducing commit(s) (\ie the only ones we considered) and other bug fixes.


\section{Conclusion} \label{sec:conclusion}

When an algorithm like SZZ becomes so prominent in software engineering research, it is more than just necessary to explore ways to ameliorate its performance. Still, it is crucial to create a platform that allows for a sound and fair comparison of any new variant.

Our goal was to create such a platform, exemplified in a publicly available and extensible oracle of multiple and documented datasets, together with open source re-implementations of a considerable number of variants.

Moreover, as we used our oracle to compare the variants and check our re-implementation validity, we came up with several concrete improvements to the existing SZZ variants.

Given the pivotal role of SZZ for various research endeavors, for example, in the context of defect analysis and prediction, and the whole field of MSR (mining software repositories), we believe our work can set the stage for numerous and, above all, comparable ameliorations of the seminal SZZ algorithm.



\section*{Acknowledgment}
This project has received funding from the European Research Council (ERC) under the European Union's Horizon 2020 research and innovation programme (grant agreement No. 851720). We are grateful for the support by the Swiss National Science foundation (SNF)  and JSPS (Project ``SENSOR'').

\balance
\bibliography{main}
\bibliographystyle{IEEEtran}

\end{document}